\documentclass[%
 aip,
 amsmath,amssymb,
 reprint,%
]{revtex4-1}

\usepackage{graphicx}
\usepackage{dcolumn}
\usepackage{bm}
\usepackage[utf8]{inputenc}
\usepackage[T1]{fontenc}
\usepackage{mathptmx}
\usepackage{etoolbox}
\usepackage{braket}   

\makeatletter
\def\@email#1#2{%
 \endgroup
 \patchcmd{\titleblock@produce}
  {\frontmatter@RRAPformat}
  {\frontmatter@RRAPformat{\produce@RRAP{*#1\href{mailto:#2}{#2}}}\frontmatter@RRAPformat}
  {}{}
}%
\makeatother
\renewcommand{\thefootnote}{\textsuperscript{\dag}}
\begin{document}

\preprint{AIP/123-QED}

\title{Magnetic field dependence of $V_B^-$\ \text{Defects} in hexagonal boron nitride }
\author{Mulin Zheng}
\thanks{These authors contributed equally to this work.}
\affiliation{Institute of Fundamental and Frontier Sciences, University of Electronic Science and Technology of China, Chengdu 610054, China}
\affiliation{James Watt School of Engineering, University of Glasgow, G12 8QQ, United Kingdom}

\author{Shizhuo Ale}
\thanks{These authors contributed equally to this work.}
\affiliation{Institute of Fundamental and Frontier Sciences, University of Electronic Science and Technology of China, Chengdu 610054, China}
\affiliation{James Watt School of Engineering, University of Glasgow, G12 8QQ, United Kingdom}

\author{Peiqin Chen}
\affiliation{Institute of Fundamental and Frontier Sciences, University of Electronic Science and Technology of China, Chengdu 610054, China}

\author{Jingpu Tu}
\affiliation{Institute of Fundamental and Frontier Sciences, University of Electronic Science and Technology of China, Chengdu 610054, China}

\author{Qiang Zhou}
\affiliation{Institute of Fundamental and Frontier Sciences, University of Electronic Science and Technology of China, Chengdu 610054, China}
\affiliation{Key Laboratory of Quantum Physics and Photonic Quantum Information, Ministry of Education,
University of Electronic Science and Technology of China, Chengdu 611731, China}

\author{Haizhi song}
\affiliation{Institute of Fundamental and Frontier Sciences, University of Electronic Science and Technology of China, Chengdu 610054, China}
\affiliation{Southwest Institute of Technical Physics, Chengdu 610041, China}

\author{You Wang}
\affiliation{Institute of Fundamental and Frontier Sciences, University of Electronic Science and Technology of China, Chengdu 610054, China}
\affiliation{Southwest Institute of Technical Physics, Chengdu 610041, China}

\author{Junfeng Wang\textsuperscript{*}}
\email{jfwang@scu.edu.cn}
\affiliation{College of Physics, Sichuan University, Chengdu 610065, China}

\author{Guangcan Guo}
\affiliation{Institute of Fundamental and Frontier Sciences, University of Electronic Science and Technology of China,  Chengdu 610054, China}
\affiliation{CAS Key Laboratory of Quantum Information, University of Science and Technology of China, Hefei 230026, China}

\author{Guangwei Deng\textsuperscript{*}}
\email{gwdeng@uestc.edu.cn}
\affiliation{Institute of Fundamental and Frontier Sciences, University of Electronic Science and Technology of China, Chengdu 610054, China}
\affiliation{Key Laboratory of Quantum Physics and Photonic Quantum Information, Ministry of Education, University of Electronic Science and Technology of China, Chengdu 611731, China}
\affiliation{CAS Key Laboratory of Quantum Information, University of Science and Technology of China, Hefei 230026, China}
\affiliation{Institute of Electronics and Information Industry Technology of Kash, Kash 844000, China}

\date{\today}

\begin{abstract}
The interface with spin defects in hexagonal boron  nitride has recently become a promising platform and has shown great potential in a wide range of quantum technologies. Varieties of spin properties of $V_B^-$ defects in hexagonal boron nitride (hBN) have been researched widely and deeply, like their structure and coherent control. However, little is known about the influence of off-axis magnetic fields on the coherence properties of $V_B^-$ defects in hBN. Here, by using the optically detected magnetic resonance (ODMR) spectroscopy, we systematically investigated the variations in ODMR resonance frequencies under different transverse and longitudinal external magnetic field, respectively. In addition, we measured the ODMR spectra under off-axis magnetic fields of constant strength but various angles, and observed that the splitting of the resonance frequencies decreases as the angle increases, aligning with our theoretical calculation based on the Hamiltonian, from which we come up with a solution of detecting the off-axis magnetic field angle. Through Rabi oscillation measurements, we found that the off-axis magnetic field suppresses the spin coherence time. These results are crucial for optimizing $V_B^-$ defects in hBN, establishing their significance as robust quantum sensors for quantum information processing and magnetic sensing in varied environments.
\end{abstract}

\maketitle

Spin defects in solid-state systems have made significant contributions to the field of quantum information applications, particularly in quantum metrology.\cite{Awschalom2018,Gao2015,Childress2006,Doherty2013,Evans2018,Zhang2020,Koehl2011,Widmann2015} Among these, nitrogen-vacancy (NV) centers in diamond and divacancy centers in silicon carbide (SiC) have established mature applications.\cite{Koehl2011,Riedel2012} In recent years, spin defects in van der Waals materials have garnered considerable attention due to their unique material structures, ultra-close proximity to the measured samples, and bandgap widths comparable to those of diamond.\\
Hexagonal boron nitride (hBN) is a common two-dimensional material with unique properties such as a wide bandgap, high thermal conductivity, and excellent chemical and mechanical stability, making it highly suitable for various sensing conditions.\cite{Cassabois2016,Kianinia2017,Jungwirth2016,Xue2018} Consequently, spin defects in hBN are among the most promising candidates for quantum sensing.\cite{Novoselov2016} While the spin properties of the $V_B^-$defect centers in hBN have been extensively studied, little is known about their spin properties in relation to off-axis magnetic field angles. This understanding is crucial for a comprehensive grasp of their spin characteristics. Additionally, ODMR signals associated with off-axis magnetic fields will aid in the development of magnetic field sensing and imaging based on $V_B^-$defect centers.\\
In this work, we investigate the spin properties of $V_B^-$defects at room temperature with respect to the off-axis magnetic field angle, focusing on the effects on ODMR resonance frequency and the coherence time as the off-axis magnetic field angle $\theta$ changes. As $\theta$ increases, the resonance frequencies of $m_s=0\rightarrow-1$ and $m_s=0\rightarrow+1$ converge towards the middle, consistent with experimental results observed in SiC.\cite{Yan2023} Furthermore, we examine the changes in coherence time with varying $\theta$ under the same magnetic field magnitude. Exploring the dependency of $V_B^-$ defects on magnetic field angles not only deepens our understanding of their spin properties but also provides valuable experimental and theoretical support for the design of future hBN-based quantum sensors. For instance, using the detected ODMR resonance frequency and the Hamiltonian of $V_B^-$ defect, we successfully fitted a random off-axis magnetic field angle, which turned out to be $68^\circ$.\\
Based on these findings, we can optimize the sensitivity and accuracy of the sensors, thereby advancing the development and practical application of quantum sensing technology. This work will facilitate the development of high-performance quantum sensors with applications in biomedical imaging, geological exploration, and fundamental physics research, showcasing broad application prospects and significant scientific importance.\\
In this study, hexagonal boron nitride (hBN) samples with spin defects are prepared via ion implantation. Bulk hBN crystals, sourced from Graphene Supermarket, are exfoliated into few-layer flakes using adhesive tape. The samples used in our measurements are approximately 100 nm thick. These flakes are initially adhered to a polydimethylsiloxane (PDMS) substrate. By pressing the PDMS against a silicon dioxide (SiO$_2$) target substrate and subsequently separating them, a portion of the hBN sample is successfully transferred onto the SiO$_2$ substrate. Nitrogen ion implantation is performed at an energy of 30 keV with a dosage of $10^{14} \, \text{cm}^{-1}$, distributed within 100 nm.\\
A home-made confocal microscope system is employed for fluorescence collection and coherent manipulation, incorporating both optical and microwave components (as illustrated in Figure 1(c)). For magnetic field control, a home-made adjustable rotary stage is used. Continuous wave 532 nm laser light is provided by a 532 nm laser (MGL-III-532nm-100mW, Changchun New Industries Optoelectronics Tech.).This laser light is modulated by an acousto-optic modulator (AOMO 3200-126, Gooch \& Housego), passes through a 550 nm short-pass filter (FESH0550, Thorlabs), and focuses onto the sample using an objective lens with a numerical aperture (N.A.) of 0.9. The fluorescence is collected through the same objective lens. Considering the emission spectrum of hBN, a 650 nm long-pass filter (FELH0650, Thorlabs) is employed to optimize fluorescence collection efficiency. The collected fluorescence is transmitted through a multimode fiber with an FC connector to a photodetector (OE-200-SI, Femto), where it is converted into an electrical signal. This signal is processed by a lock-in amplifier (MFLI, Zurich Instruments) to extract spin information and then transmitted to a data acquisition card (USB-6351, NI) for analysis.\\
In the microwave system, microwave signals generated by an integrated microwave source (SSG-6000RC, Mini-Circuits) are modulated by a microwave switch (ZASWA-2-50DRA+, Mini-Circuits), controlled by a pulse generator (PBESR-PRO-500, Spincore). These modulated signals are amplified by a 45 dB amplifier (UWBPAEG6G-30W, RFUWB) and directed onto the sample using a coplanar waveguide antenna (Figure 1(c)).\\
\begin{figure}[h]
    \centering
    \includegraphics[width=0.5\textwidth]{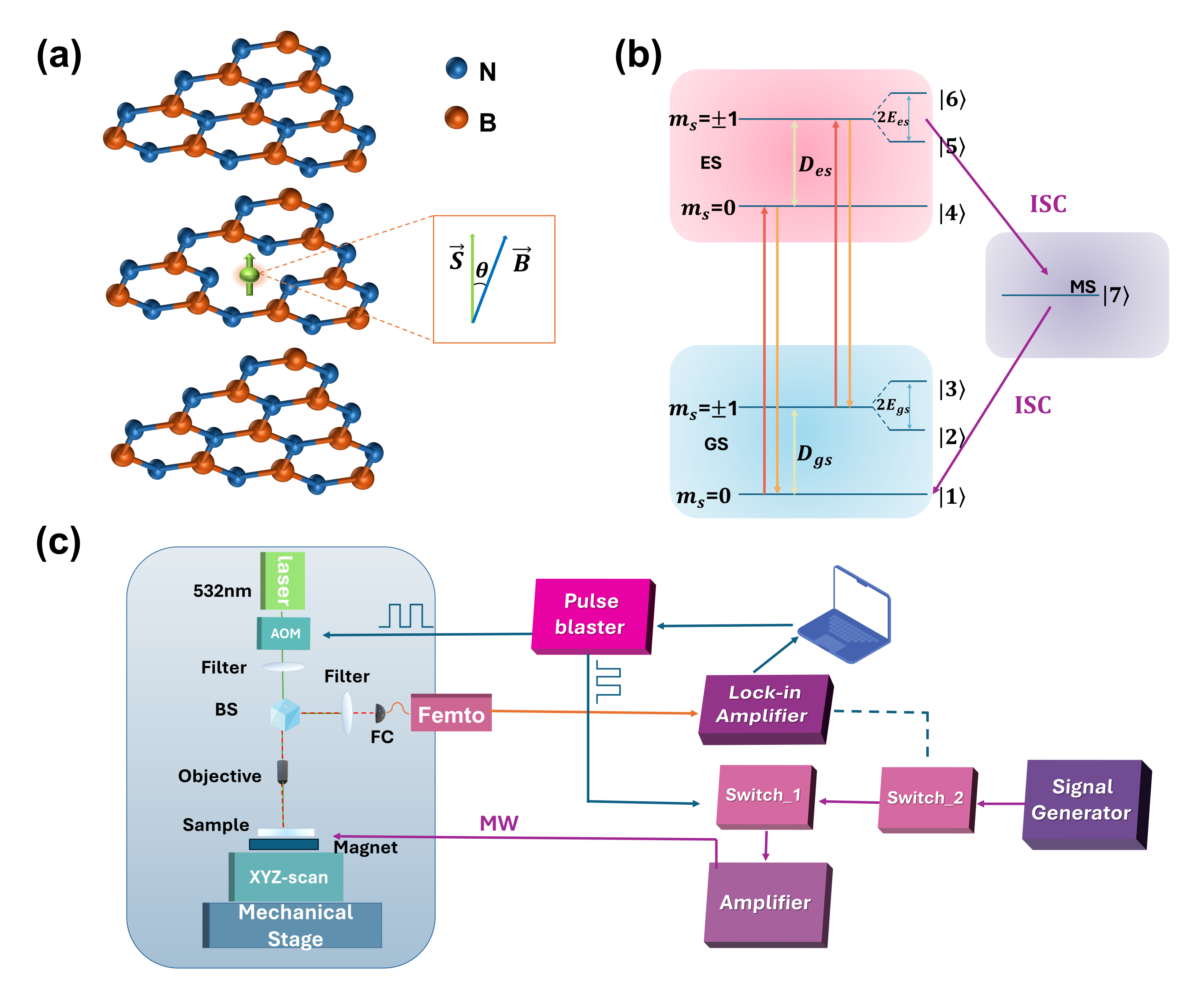}
    \caption{\textbf{Schematic diagrams of the system.} (a) Schematic diagram of the $V_B^-$ defect structure. The blue spheres represent nitrogen atoms, the orange spheres represent boron atoms. Inset: the green arrow indicates the spin direction (i.e., the symmetric c-axis), and the blue arrow indicates the off-axis magnetic field direction. The angle between the external magnetic field and the c-axis is $\theta$. (b) Schematic diagram of the energy level structure. The red, purple, and orange arrows represent different transition pathways. $D_{es}$ and $D_{gs}$ are the splitting parameters in the spin Hamiltonian. (c) Schematic diagram of the confocal system experimental setup. This includes the optical cage system (blue part) and the microwave system. BS: Dichroic mirror; FC: Fiber coupler; AOM: Acousto-optic modulator; Femto: Fast photodetector.}
    \label{fig:1}
\end{figure}
In the injected hBN samples, boron vacancy ensembles are distributed throughout the sample and exhibit stable fluorescence intensity at room temperature. As shown in Figure 1(a), we apply an off-axis magnetic field $\vec{B}$ and denote the magnetic field angle of the off-axis field by the angle $\theta$ between the external magnetic field direction  $\vec{B}$ and the spin direction $\vec{S}$. Based on theoretical studies\cite{Huang2012,Abdi2018,Reimers2020,Sajid2020,Ivady2020,Barcza2021} and detailed characterizations of the energy levels of the excited state \cite{Mathur2022,Baber2022,Yu2022,Zhao2021}, we have gained a clear understanding of the energy level system of the $V_B^-$ defect center. As depicted in Figure 1(b), the triplet ground state (triplet GS) transitions to the triplet excited state (triplet ES) by absorbing energy and then returns to the triplet ground state through radiative recombination. This process also includes non-radiative intersystem crossing (ISC) from the excited state to the ground state. The $V_B^-$  defect is continuously polarized to the $m_s=0$ state by ISC under continuous laser illumination. When the applied microwave frequency resonates with the energy difference between $m_s=0\rightarrow-1$ or $m_s=0\rightarrow+1$, electrons are pumped from the $m_s=0$ state to the $m_s=\pm1$ state, resulting in a decrease in fluorescence intensity. First, we characterized the dependence of the ODMR resonance frequency of $V_B^-$ on the longitude and transverse external magnetic field (Figure 2).
\begin{figure}[htbp]
    \centering
    \includegraphics[width=0.5\textwidth]{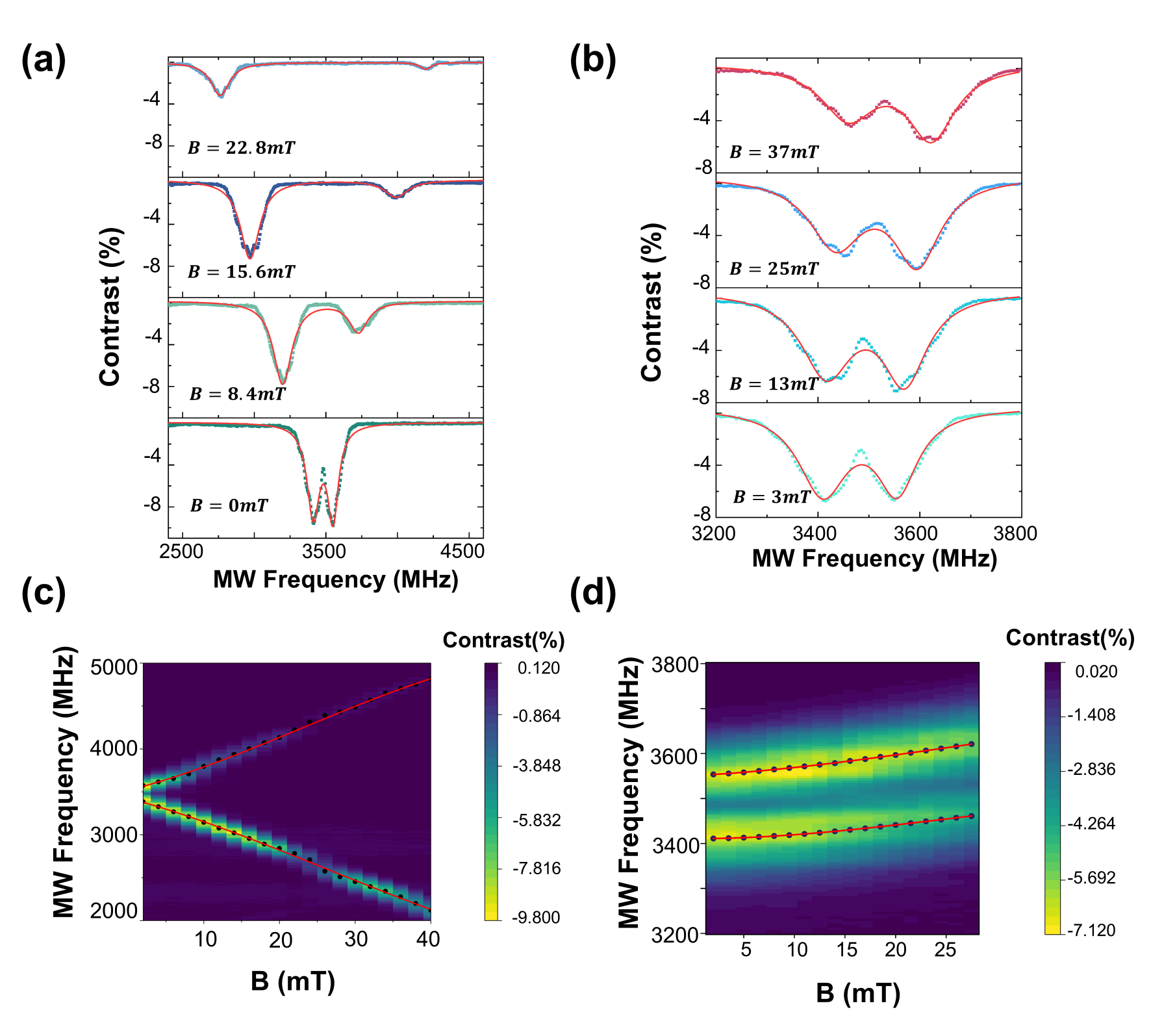}
    \caption{ \textbf{Magnetic field-dependent continuous ODMR spectra.} (a) and (b) ODMR spectra at different magnetic field magnitudes with $\theta=0^\circ$ (parallel to the \(c\)-axis) and $\theta=90^\circ$ (perpendicular to the \(c\)-axis). The red solid lines are the fitted curves, and the colored dots are the experimental data points. (c) and (d) Dependence of CW ODMR on the same magnetic field magnitude at $\theta=0^\circ$ and $\theta=90^\circ$. The black dotted lines are the experimental data, and the red solid lines are the fitted curves.}
    \label{fig:example}
\end{figure}
\addcontentsline{lof}{figure}{\protect\numberline{\thefigure}{ODMR at different magnetic field magnitudes.}} According to the $V_B^-$ spin ground state Hamiltonian \cite{Schirhagl2014}:
\begin{equation}
H = D \left( S_z^2 + \frac{2}{3}I \right) + E \left(S_x^2 + S_y^2\right) + \gamma_e \vec{B} \cdot \vec{S}
\end{equation}\\
where \( D \) is the zero-field splitting parameter, \( E \) is the strain splitting parameter, and \( \gamma_e \) is the gyromagnetic ratio of the electron spin. Here, we use the values \( D_{gs} \sim 3.49\, \text{GHz} \), \( E_{gs} \sim 50\, \text{MHz} \), and \( \gamma_e \sim 2.8\, \text{MHz/Gauss} \). \( \vec{S} \) is the spin-1 vector operator, \( \vec{B} \) is the external vector magnetic field, and \( I \) is the identity matrix. The \( \frac{2}{3} I \) term is used to maintain the symmetry and energy scale consistency of the Hamiltonian. The third term in the equation, known as the Zeeman interaction term, describes the effect of the external local field on the system.
Under an off-axis external magnetic field, we can calculate the eigenstates of the ground state \(\ket{1}\) to \(\ket{7}\). These can be represented as linear combinations of the ground state \(\ket{0},\ket{-1},\ket{+1}\), the excited state\(\ket{0},\ket{-1},\ket{+1}\), and the metastable state, respectively \cite{Tetienne2012}:\\
\begin{equation}
    \ket{i} = \sum_{j=1}^{7} \alpha_{ij}(B) \ket{j^0}, \quad i = 1, 2, 3, 4, 5, 6, 7
\end{equation}
Based on above understanding, we can incorporate the transverse and longitudinal components of the magnetic field into the Zeeman interaction term for calculation. The total Hamiltonian can theoretically predict the ODMR resonance frequency as a function of the magnetic field angle $\theta$ (see Supplementary Information). As shown in Figure 2(c), the red solid line represents the fit based on the Hamiltonian, which matches the experimental results (black dotted line), i.e., at $\theta=0^\circ$, the resonance frequencies of $m_s=0 \rightarrow -1$ and $m_s=0 \rightarrow +1$ split away as the magnetic field increasing. To further investigate the response of ODMR to the transverse magnetic field (i.e., perpendicular to the c-axis), we examined the ODMR spectra at $\theta=90^\circ$ under different magnetic field magnitudes. As shown in Figures 2(b) and 2(d), at $\theta=90^\circ$, the resonance frequencies corresponding to the two ODMR dips increase with increasing magnetic field magnitude,and the splitting remains almost the same. According to the discussion in the Supplementary Information, we attribute this to the Zeeman effect. The Zeeman interaction term in the Hamiltonian is affected by the transverse magnetic field, leading to changes in $B_x S_x$, and consequently causing variations in the resonance frequency (see Supplementary Information).\\
Based on the above results, we conclude that variations in the off-axis magnetic field angle, even when the magnetic field strength remains constant, can induce changes in the ODMR signal of the \(V_B^-\) spin defects. Therefore, by keeping the magnetic field strength fixed at 16.4 mT and rotating the off-axis magnetic field angle, we were able to establish the relationship between ODMR and the off-axis magnetic field angle. As shown in Figs. 3(a) and 3(c), the splitting of ODMR resonance frequencies decreases as the off-axis magnetic field angle $\theta$ increases. This result is consistent with the findings in the divacancy system in SiC.~\cite{Shin2013} Based on our calculations presented in the supplementary materials, we extract the angle $\theta$ through data fitting with the eigenenergies of the ground-state spin Hamiltonian, \(H_{gs}\). From the ODMR spectra, two dips are observed in the PL signal, corresponding to the electron spin transitions \(|0\rangle \rightarrow |+1\rangle\) and \(|0\rangle \rightarrow |-1\rangle\).\\

\begin{figure}[htbp]
    \centering
    \includegraphics[width=0.5\textwidth,keepaspectratio]{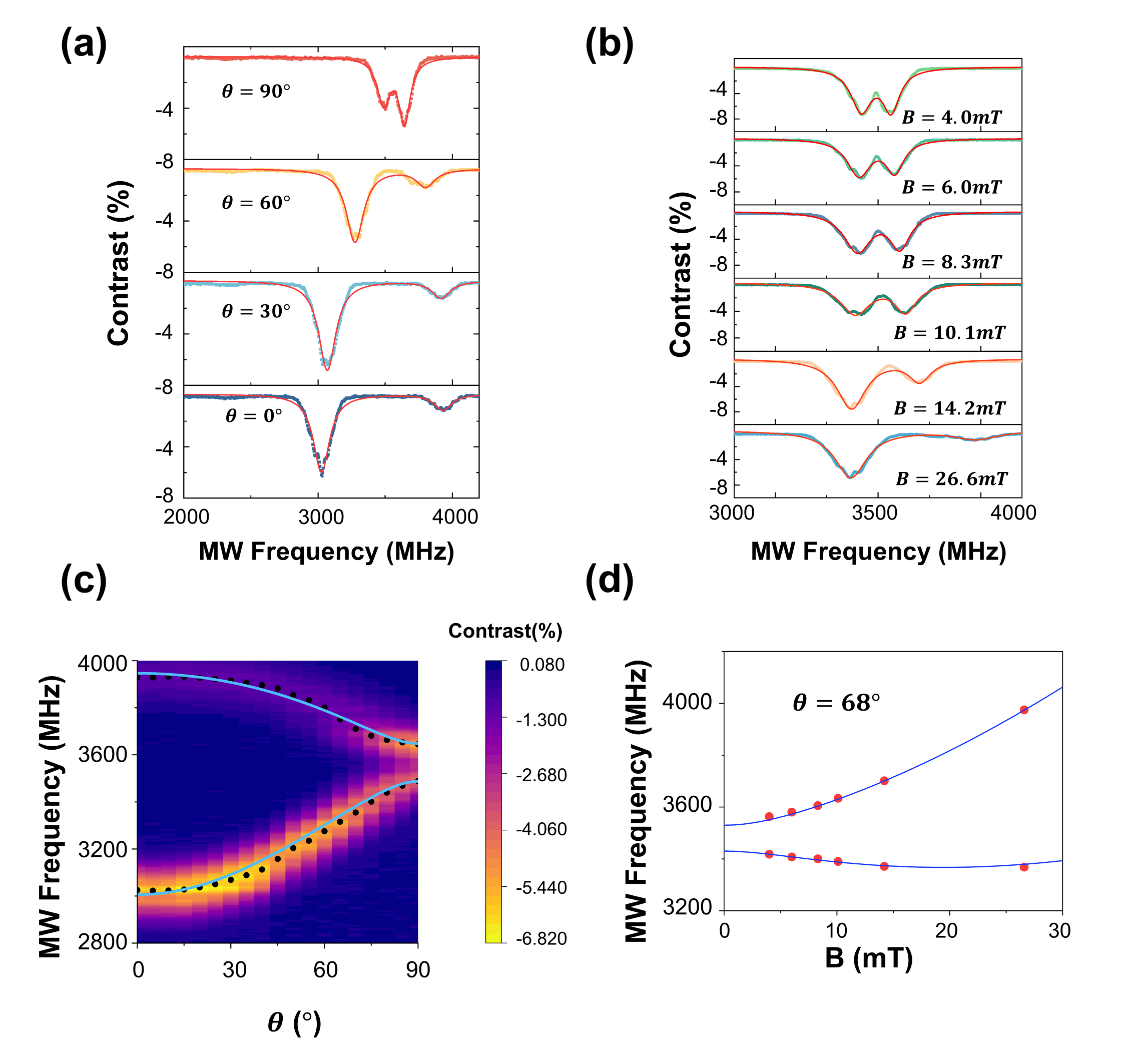}
    \caption{\textbf{Demonstration of off-axis magnetic field angle fitting.} (a) ODMR spectra at a magnetic field magnitude of 16.7 mT for different off-axis magnetic field angles. The red solid lines are the fitted curves, and the colored dots are the experimental data points. (b) ODMR spectra under varying external magnetic field magnitudes with a fixed off-axis magnetic field angle. Colored dots represent the measured data, while the red solid lines represent the fitted curves. The ODMR spectra were recorded at various magnetic field magnitudes. (c) Dependence of CW ODMR on the off-axis magnetic field angle at room temperature with B=16.7 mT. The black dotted line represents the experimental data, and the blue solid line is the fitted curve. (d) Determination of the off-axis magnetic field angle from the relationship between the magnetic field magnitude and resonance frequency. Red dots represent the measured data, and the blue solid line corresponds to the data fitting with the eigenenergies of \(H_{gs}\). The angle deduced from the fit is $\theta=68^\circ$.}
    \label{fig:example}
\end{figure} 

By measuring the ODMR spectra of \(V_B^-\) spin defects at varying magnetic field magnitude but fixed magnetic field orientation (Fig.3(b)), we obtained the resonance frequencies of the ODMR as a function of the magnetic field magnitude. This allows us to fit the off-axis magnetic field angle basing on the Hamitonian, as shown in Fig.3(c), with the measured angle between the off-axis magnetic field and the spin direction \(c\)-axis being $68^\circ$. This allows us to use the ODMR spectra at different magnetic field magnitude to determine the off-axis magnetic field angle.\\

\begin{figure}[htbp]
    \centering
    \includegraphics[width=\linewidth]{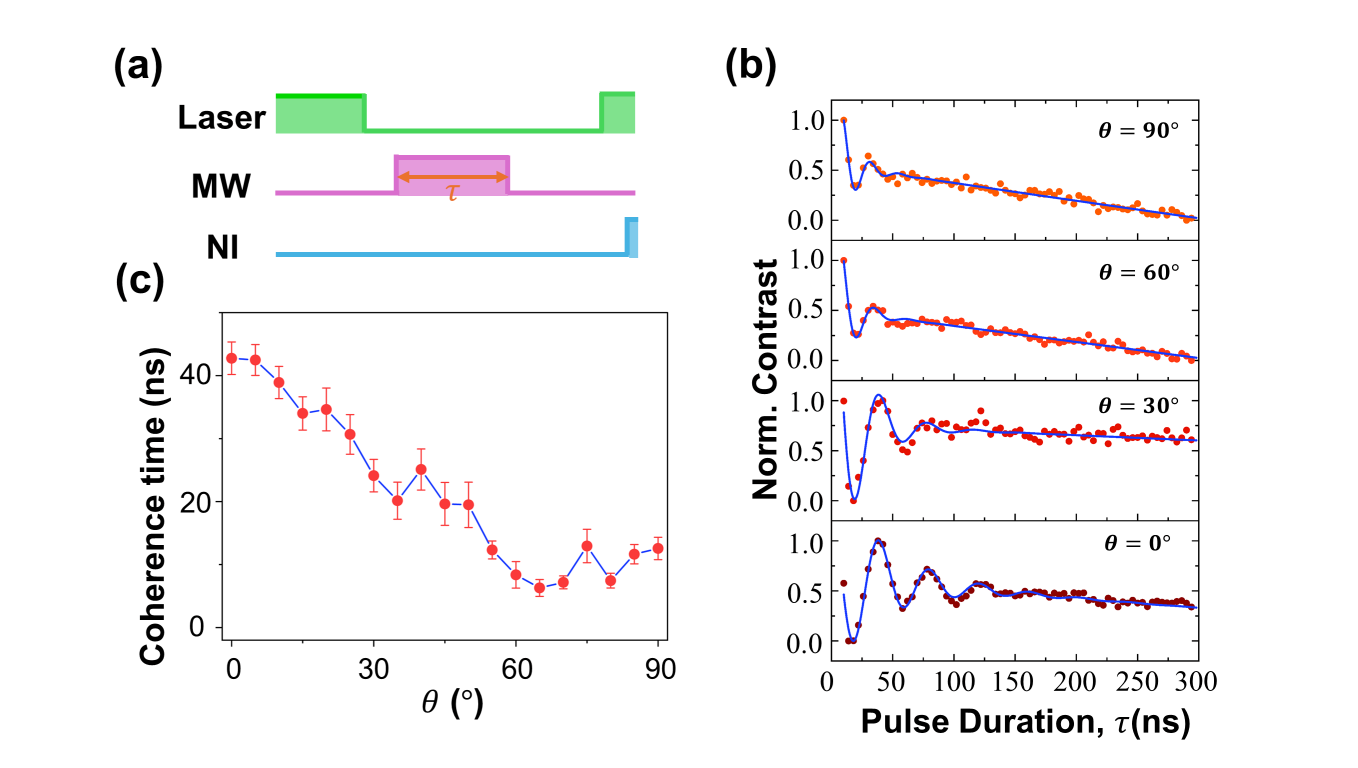}
    \caption{\textbf{Response of coherence time to the off-axis magnetic field angle $\theta$.} (a) Pulse sequence for Rabi oscillations. (b) Rabi oscillations at magnetic field angles of $0^\circ$, $30^\circ$, $60^\circ$, and $90^\circ$. The blue solid lines are the fitted curves, and the red dots are the experimental data points. (c) Dependence of coherence time on the magnetic field angle.}
    \label{fig:example}
\end{figure} We then fit the data using a damped oscillation function:\cite{Liu2022}
\begin{equation}
f(\tau) = ae^{-\tau/T_a} \cos(2\pi f \tau + \varphi) + be^{-\tau/T_b} + c
\end{equation}

Off-axis magnetic fields also have a certain impact on the coherence of the system.\cite{Stanwix2010,Mittiga2018} To investigate the influence of off-axis magnetic fields on the coherence of the spin system in detail, we measured the Rabi oscillations of the $V_B^-$ defect with different magnetic field angles $\theta$ using the pulse sequence shown in Figure 4(a), and the results are in Figure 4(b).

Here, $T_a$ is related to the system coherence and can reflect the coherence time of the $V_B^-$ defect. \cite{Stanwix2010} Based on this, we measured the Rabi oscillations of the $V_B^-$ defect under an external static magnetic field with different magnetic field angles $\theta$. Figure 4(b) shows the Rabi oscillations at magnetic field angles of $0^\circ$, $30^\circ$, $60^\circ$, and $90^\circ$. The red dots represent the experimental data, indicating that the intensity of the Rabi oscillations decreases as the angle increases. The blue solid line represents the fitted curve. From the fitting, we studied the coherence time as a function of the magnetic field angle (Figure 4(c)). The coherence time decreases with the increase in the off-axis magnetic field angle. When the direction of the external magnetic field aligns with the spin direction of the $V_B^-$ defect, the maximum intrinsic coherence time of 42.76 ns can be obtained.\\

The main reason for the decrease in coherence time with increasing off-axis magnetic field angle is due to the spin bath precession that depends on the magnetic field and the impact of the magnetic field angle on the hyperfine interaction between the electron and nuclear spins. On one hand, changes in the magnetic field angle alter the Larmor frequency of the $^{14}$N nuclear spin precession. \cite{Maze2008,Shin2013,Gao2022} As the magnetic field angle $\theta$ increases, the variation in the nuclear spin's precession frequency components becomes larger. The enhanced decoherence effect of the spin bath on the electron spin results in a shorter coherence time. \cite{Maurer2012,Zhao2011} On the other hand, when the direction of the external magnetic field changes, the alteration in the hyperfine interaction tensor $A(\theta, \varphi)$ changes the form and strength of the interaction between the electron spin and the nuclear spin. When $\theta=0^\circ$, the magnetic field direction aligns with the system's symmetry axis, and the hyperfine interaction primarily acts in one direction. As $\theta$ increases, the hyperfine interaction distributes across multiple directions, making it more complex. This accelerates the decoherence of the $V_B^-$ spin, leading to a further reduction in coherence time. \cite{Takahashi2008,Witzel2006, Smeltzer2011}\\


In summary, we studied the impact of off-axis magnetic fields on the coherence properties of $V_B^-$ defects in hBN. Using ODMR spectroscopy, we found that the splitting of the two resonance frequencies decreases with increasing magnetic field angle $\theta$, consistent with our theoretical model. By measuring the ODMR spectra under different magnetic field magnitude, the variation in resonance frequency allows us to fit the angle of the off-axis magnetic field. The contrast of the ODMR signal also varies with $\theta$, exhibiting different behaviors under longitudinal and transverse magnetic fields. Rabi oscillation measurements indicated that the coherence time significantly shortens as $\theta$ increases. This reduction in coherence time is attributed to enhanced decoherence caused by magnetic field-dependent spin bath precession and complex hyperfine interactions. These findings provide important insights for optimizing the application of $V_B^-$ defects in hBN for quantum information processing and magnetic sensing. They highlight the potential of $V_B^-$ defects as reliable quantum sensors in various magnetic environments.
\section*{Acknowledgements}
This work was supported by National Key Research and Development Program of China (No. 2022YFA1405900), National Natural Science Foundation of China
(No. 12074058), Sichuan Science and Technology Program (Nos. 2024YFHZ0372).
\section*{Data Availability Statement}
The data that support the findings of this study are available from the corresponding author upon reasonable request.

\nocite{*}
\bibliography{aipsamp}

\end{document}